\documentclass[11pt]{article}
\usepackage{amssymb,amsmath,epsfig}

\begin{document}
\thispagestyle{empty}
\null \hfill AEI-2005-015\\

\vskip 1 cm

\begin{center}
{\Large \bf A two-loop test for the factorised S-matrix of planar
${\cal N}=4$} \\

\vskip 1truecm
{\bf B. Eden} \\

\vskip 0.5 cm

{\it Max-Planck-Institut f\"ur Gravitationsphysik,
Albert-Einstein-Institut, \\
Am M\"uhlenberg 1, D-14476 Golm, Germany} \\

\end{center}

\vskip 2 cm

\Large \centerline{\bf Abstract} \normalsize

We check the recently proposed higher loop Bethe-ansatz for the $sl(2)$
sector of ${\cal N}=4$ at two loops by a direct perturbative calculation
using ${\cal N}=2$ superfields in supersymmetric dimensional reduction.
Our method can in principle address operators built from many
elementary fields and of arbitrary twist. We work out in detail the spin
three, twist three operator mixing problem at leading order in $N$ and find
agreement with the prediction based on integrability.

\newpage

\section{Introduction}

The original version of the AdS$_5$-CFT$_4$ correspondence
is a weak-strong duality so that a direct comparison
of string theory with gauge theory quantities is hardly possible in the
perturbative regime. In a special limit of the geometry this difficulty may
have been overcome by the BMN-proposal \cite{Berenstein:2002jq}.
The exact leading $N$ orthogonalisation of the relevant set of operators has
been connected to integrable spin chains, thus opening the possibility of
analysing the large $N$ mixing in ${\cal N}=4$ through the Bethe-ansatz
\cite{Minahan:2002ve}. The spin chain Hamiltonian can be identified with the
dilatation operator of the planar CFT \cite{dilop}, a powerful tool that has
successfully been used e.g. to compute one- and two-loop scaling dimensions
\cite{Beisert:2003ea} of the gauge theory equivalent of a class of string
theory solitons with large angular momentum \cite{frotse}.

On the field theory side, dilatation operator and Bethe-ansatz are known for
the full theory at one-loop \cite{oneloop}, while the dilatation operator has
been worked out up to three loops for an $su(2|3)$ subsector \cite{su23}, and
if the subalgebra is further restricted to $su(2)$ also to four loops
\cite{su24l} and five loops \cite{su2al}. For the $su(2)$ sector a Bethe-ansatz
up to three loops was formulated in \cite{stauserb} and there is even an
all-loops conjecture based on a new long-range spin chain \cite{su2al}.

A step towards the quantisation of AdS$_5\times$S$_5$ string theory has been
made in \cite{callan} by taking into account the lowest term in a 
perturbation series in the curvature around the $pp$-wave
background underlying the BMN construction. Data from here as well as from the
soliton calculations has been incorporated into a Bethe-ansatz on the string
side \cite{stringbethe}. The ansatz deviates from the gauge theory conjecture
starting at three loop order in the gauge coupling, where there are extra
scattering terms, although these are expressed in terms of quantities known
from the gauge theory \cite{su2al}. The three-loop disagreement of the
Bethe-ansaetze accurately reproduces the gauge theory/string theory mismatches
found up to now \cite{stauserb,callan}. Despite of this, the integrable
structures on the two sides of the conjecture are very similar: one may for
example match the higher charges \cite{charges} at one and two loops.
Furthermore, the spectrum of the
classical string in the $su(2)$ sector was derived from a set of integral
Bethe-equations in \cite{KMMZ}, the equivalent of the thermodynamic limit of
the discrete Bethe equations in the gauge theory picture. The idea behind the
string Bethe-ansatz was in fact to look for an appropriate discretisation of
the continuum equations. The ansatz has the feature of ``factorised
scattering'',~i.e. that the Bethe equations involve a product of two-site
interactions termed $S$-matrices.

The article \cite{new} shifts the attention entirely to the $S$-matrix. It is
observed that the energy eigenvalues found in \cite{callan} for an $su(1|1)$
part of the $su(2|3)$ sector and in \cite{kaszar} for the disjoint $sl(2)$
sector are compatible with factorised scattering. A relation between the string
theory $S$-matrices in the three sectors is transferred to gauge theory,
yielding a two- and three-loop Bethe-ansatz in the $sl(2)$ sector. The
associated spectrum is checked for the twist two operators, whose
anomalous dimensions up to three loops were extracted in \cite{lipatov} from
an outstanding QCD calculation \cite{moch}.

The purpose of the present paper is to test the two-loop anomalous dimensions
(i.e. order $g^4$) given in \cite{new} for the twist three, spin three
operators in the $sl(2)$ sector against a direct calculation of two-point
functions based on ${\cal N}=2$ superfields \cite{HSS} and regularisation by
supersymmetric dimensional reduction (SSDR) \cite{DimRed}.\footnote{Superspace
calculations package ordinary Feynman graphs into supergraphs.
In many situations this leads to a significant decrease of the number of
diagrams. We prefer the ${\cal N}=2$ over the ${\cal N}=1$ formulation
because the extra supersymmetry helps to suppress even more graphs.} The result
strongly vindicates the $S$-matrix conjecture of \cite{new},
since at least in the dilatation operator picture there is the issue of
``wrapping'',~i.e. how the general interaction is modified when there are
only few sites in the spin chain. We emphasize that our method is applicable
beyond twist two; barring for computing limitations we can handle operators
built from many elementary fields and carrying higher spins.

We encounter Feynman graphs with up to four loop
integrations in momentum space.\footnote{In the literature the
$O(g^2),O(g^4)$ parts of the scaling dimension are usually called one-loop and
two-loop anomalous dimension. Nevertheless, when the first two anomalous
dimensions are computed from two-point functions the momentum space
Feynman integrals have two loops at order $g^2$ and up to four at $g^4$.}
The only four-loop diagram is of a type that we call
``BPS-like'' since the same supergraphs without extra partial derivatives
on the outer legs occur in the two-point functions of 1/2 BPS operators.
For the two- and three-loop diagrams of this type we verify finiteness
with and without extra partial derivatives using the cutting edge \emph{Mincer}
algorithm \cite{Mincer}. By analogy, we assume that also the four-loop
BPS-like supergraph does not contribute. On the other hand, all the singular
integrals have maximally three loop integrations in momentum space so that
they can be evaluated exactly.

Note that the \emph{Mincer} programme itself calculates massless two-point
integrals with two or three loop integrations and arbitrary \emph{scalar}
numerators. We reconstruct integrals with open indices from their projections
with the total ingoing momentum.

Future work aims at the construction of the planar two-loop dilatation
operator for the $sl(2)$ sector, which would allow one to further analyse 
the recently discovered one-loop gauge/string theory discrepancies in the
$sl(2)$ sector \cite{tseyt}.

\section{The mixing problem}

We consider the leading $N$ mixing of the operators
\begin{equation}
\{s_1,s_2,s_3\} \, = \, Tr \bigl( (D_z^{s_1} Z) (D_z^{s_2} Z) (D_z^{s_3} Z)
\bigr) \, . \label{opdef}
\end{equation}
Here $Z$ is a complex scalar field of the ${\cal N} = 4$ SYM theory with
$SU(N)$ gauge group and $D_\mu \, = \, \partial_\mu + \, i \, g \, A_\mu$
is the Yang-Mills covariantised space-time derivative. The operators carry
spin $s = s_1 + s_2 + s_3$,~i.e. the Lorentz indices are symmetrised and
traceless, which can be made automatic by projecting all indices onto
$z = x_1 + i x_2$ as indicated in (\ref{opdef}).

The theory is regularised by SSDR (supersymmetric dimensional reduction)
\cite{DimRed} in $x$-space. This scheme is related to dimensional
regularisation, where the scalar propagator is
\begin{equation}
\langle Z(1) \bar Z(2) \rangle \, = \, \frac{c_0}{x_{12}^2} (\mu x_{12}^2)
^\epsilon \, , \qquad c_0 \, = \, - \frac{1}{4 \pi^2} \, , \qquad
\square_1 \langle Z(1) \bar Z(2) \rangle \, = \, \delta(x_{12}) \, .
\label{ssdr}
\end{equation}
(The normalisation change involving non-integer powers of $\pi$ and $\zeta(2)$
is hidden in the mass scale. We will not usually write $\mu$, since it can
easily be reinstated.) On dimensional grounds, Feynman-graphs must come with
the fractional powers
\begin{equation}
g^0 : \, (x_{12}^2)^{3 \epsilon}, \, \qquad g^2 : \, (x_{12}^2)
^{4 \epsilon}, \, \qquad g^4 : \, (x_{12}^2)^{5 \epsilon} \, ,
\end{equation}
whose $\epsilon$-expansion yields logarithms. The conformal invariance of
${\cal N} = 4$ SYM implies that these logs exponentiate to scaling dimensions
associated with composite operators.

Such conformal properties become visible only in the renormalised theory,
i.e. after introduction of ${\cal Z}$-factors and in the limit $\epsilon
\rightarrow 0$. The set of operators has to be orthogonalised, upon which the
two-point function of a correctly renormalised primary of spin $s$ has the form
\begin{equation}
\langle P^s(1) \bar P^s(2) \rangle \, = \, \frac{c(g^2) J_{\mu_1 \nu_1}(x_{12})
\ldots J_{\mu_s \nu_s}(x_{12})}{(x_{12}^2)^{\Delta(g^2)}} \label{conf}
\end{equation}
where the sets of indices $\{\mu_1 \ldots \mu_s\}$ and $\{\nu_1 \ldots \nu_s\}$
are both individually symmetrised and made traceless \cite{osborn}. In the last
equation we have used the inversion tensor
\begin{equation}
J_{\mu \nu}(x) \, = \, \eta_{\mu \nu} - 2 \frac{x_\mu x_\nu}{x^2}
\end{equation}
to organise the terms in a simple manner. The scaling dimension
$\Delta = \Delta_0 + g^2 \gamma_1 + g^4 \gamma_2 + \ldots$ splits
into a classical part $\Delta_0$ and the ``anomalous dimension'' depending
on $g^2$. It can clearly be read off from any term in (\ref{conf}).
We find it convenient to drop all traces because this reduces the set of
graphs, see below. If desired, the full functional form of any two-point
function can be reconstructed: for primary operators we may use (\ref{conf})
while two-point functions of conformal descendants are obtained from it
by differentiation.

We are interested in the anomalous dimension up to order
$g^4$, which requires the leading and the next-to-leading term in the
$\epsilon$-expansion,~i.e.
\begin{equation}
g^0 : \, O(1),O(\epsilon), \, \qquad g^2 : \, O(\frac{1}{\epsilon}), O(1),
\, \qquad g^4 : O(\frac{1}{\epsilon^2}), O(\frac{1}{\epsilon}) \, .
\end{equation}
In particular, at $g^4$ we do not need to keep track of the finite part of the
integrals. Correspondingly, we will not eliminate finite $O(g^4)$ cross terms
between the operators. Terms of order $\epsilon$ and higher yield contact terms
when the regulator is taken off \cite{contact}. Although we are generally not
interested in the contact part we cannot discard the $O(\epsilon)$ terms in the
tree-level mixing, because the singular renormalisation factors relate them to
the subleading terms at $g^2$ and $g^4$. The renormalisation/orthogonalisation
proceeds along these lines: first, we need to choose a basis in which the
$O(1)$ tree-level mixing and the simple logarithm terms at $g^2$ are
diagonalised. In such a basis each operator picks up a ${\cal Z}$-factor of the
standard form \cite{Eden:2005ve}
\begin{equation}
{\cal Z} \, = \, 1 + g^2 \frac{\gamma_1}{2 \epsilon} + g^4 \Bigl( \frac{
\gamma_1^2} {8 \epsilon^2} + \frac{\gamma_2}{4 \epsilon} \Bigr) + O(g^6) \, ,
\label{renz}
\end{equation}
and $\gamma_1,\gamma_2$ can already be calculated. Second, in order to
fully orthogonalise we may have to subtract the other operators with
mixing coefficients of the type
\begin{equation}
{\cal B} \, = \, g^2 b_{10} + g^4 \Bigl( \frac{b_{21}}
{\epsilon} + b_{20} \Bigr) + O(g^6) \, , \label{renb}
\end{equation}
(In this letter $b_{20}$ remains undetermined because we truncate the
$\epsilon$-expansion.)
This form of the ${\cal Z}$-factors may be called ``minimal subtraction'' since
we strictly eliminate only singularities and finite cross terms. \\

At \textbf{spin 0} we have the operator
\begin{equation}
O \, = \, \{0,0,0\} \, .
\end{equation}

At \textbf{spin 1} we find its descendant
\begin{equation}
D_z O \, = \, \{1,0,0\} \, .
\end{equation}

At \textbf{spin 2} there are
\begin{eqnarray}
D_z^2 O & = & \{2,0,0\} + 2 \, \{1,1,0\} \, , \label{defK6} \\
K_6 & = & \{2,0,0\} - (2 + \epsilon) \, \{1,1,0\} \, . \nonumber 
\end{eqnarray}

The operator $O$ is a 1/2 BPS state. It is gratifying to see that its $g^2,g^4$
two-point functions do indeed not contain singular or finite pieces.
For $x_1 \neq x_2, \, \epsilon \rightarrow 0$ the loop contributions
to $\langle O \bar O \rangle$ converge to zero and thus our calculation
confirms protectedness.
Let the descendants $D_z^s O, \, s = 1,2,3$ be defined
by naive distribution of each derivative over the three sites in
the operator. All three descendants are exact null-vectors of the $g^2$ and
$g^4$ mixing.

At spin 2 there is a second operator which we must orthogonalise w.r.t.
$D_z^2 O$. The definition (\ref{defK6}) for $K_6$
is not quite in the spirit of minimal subtraction because we have used a
mixing coefficient involving a positive power of $\epsilon$. It has the
advantage that the bare $K_6$ is tree-orthogonal to $D_z^2 O$ at leading and
next-to-leading order in $\epsilon$, whereas this is guaranteed at $g^2,g^4$.
Thus with the choice (\ref{defK6}) the renormalised operator $K_{6 \,ren} \,
= \, {\cal Z}_K K_6$ decouples from $D_z^2 O$. Using a standard
${\cal Z}$-factor we find
\begin{equation}
\langle K_{6 \, ren} \bar K_{6 \, ren} \rangle \, = \, - 128 M^3 (1 + g^2 M +
\ldots) \frac{x_{12 \, z} x_{12 \, z} x_{12 \, \bar z} x_{12 \, \bar z}}
{(x_{12}^2)^{(7 + 4 g^2 M - 6 g^4 M^2 + \ldots)}} - traces \label{k6twop}
\end{equation}
with the loop-counting factor $M = N / (4 \pi^2)$. The values
\begin{equation}
\gamma_1 \, = \, 4 M \, , \qquad \gamma_2 \, = \, - 6 M^2
\end{equation}
clearly identify the operator $K_6$ as belonging to the BMN-multiplet
starting with the super-primary state $Tr(Z \Phi_I \bar \Phi^I) + Tr(Z \bar
\Phi^I \Phi_I)$ analysed in \cite{dilop,Eden:2005ve}.\footnote{The
convention used in \cite{Eden:2005ve} differs by a rescaling of the
coupling constant.}

For minimal subtraction in the tightest sense we might choose
$K_6' \, = \, \{2,0,0\} - 2 \, \{1,1,0\}$. The divergence in ${\cal Z}_K$ then
sees the $O(\epsilon)$ term in $\langle K_6' \, D_z^2 O
\rangle$ and forces a renormalisation
\begin{equation}
K_{6 \, ren} = {\cal Z}_K \, K_6' - (g^2 \frac{M}{2} + g^4 \Bigl( \frac{M^2}
{2 \epsilon} + b_{20} \Bigr) + O(g^6)) \, D_z^2 O
\end{equation}
where ${\cal Z}_K$ still has standard form. This is not multiplicative
as ${\cal Z}_K = 1 + 2 g^2 M / \epsilon + \ldots \, .$ To the order we work
there is apparently freedom in the system, the stable feature being the
anomalous dimension of $K_6$. \\

For the \textbf{spin 3} mixing problem we return to our somewhat more elegant
first scheme, namely to make the operators tree-orthogonal at leading and
next-to-leading order. This is achieved by
\begin{eqnarray}
D_z^3 O & = & \{3,0,0\} + 3 \, (\{2,1,0\} + \{1,2,0\}) + 2 \, \{1,1,1\} \, ,
\label{defspin3} \\ D_z K_6 & = & \{3,0,0\} - (1+\epsilon) \, (\{2,1,0\} +
\{1,2,0\}) - (2+\epsilon) \, \{1,1,1\} \, , \nonumber \\
V_1 & = & (2 - \frac{68 \epsilon}{35}) \, \{3,0,0\} - (9 - \frac{96 \epsilon}
{35}) \, (\{2,1,0\} + \{1,2,0\}) + (24 +
\frac{164 \epsilon}{35}) \, \{1,1,1\} \, , \nonumber \\
V_2 & = & \{2,1,0\} - \{1,2,0\} \, . \nonumber
\end{eqnarray}
The operator $V_2$ is the only one that is odd under reversal of the
three sites in the trace whereby it completely decouples at classical and
quantum level. The first two operators $D_z^3 O$ and $D_z K_6$ are defined
by naive differentiation of their spin 2 ancestors. As already mentioned,
$D_z^3 O$ is a null-vector of the loop-level mixing; in conclusion
this operator does not mix either.

One may wonder what correlation functions $D_z K_6$ and $V_1$ have w.r.t.
the conformal primary $K_6$ itself. By direct calculation
\begin{equation}
\langle K_{6 \, ren} \, D_{\bar z} \bar K_{6 \, ren} \rangle \, = \,
- 1792 M^3 (1 + \frac{11}{7} \, g^2 M + \ldots) \frac{x_{12 \, z} x_{12 \, z}
x_{12 \, \bar z} x_{12 \, \bar z} x_{12 \, \bar z}}
{(x_{12}^2)^{(8 + 4 g^2 M - 6 g^4 M^2 + \ldots)}} - traces
\end{equation}
as expected by differentiating formula (\ref{k6twop}).\footnote{Likewise, the
direct evaluation of $\langle D_z K_{6 \, ren} \, D_{\bar z} \bar K_{6 \, ren}
\rangle$ yields the expected $x$-dependence with the correct
normalisation $28672 M^3 (1 + 29/14 \, g^2 M + \ldots ) \, .$ The point may be
worth mentioning as the projection method in the graph calculation is
relatively involved.} More interesting is the question as to how
$V_1$ can be made orthogonal to $K_6$. We start with the
definition
\begin{equation}
V_{1 \, ren} \, = \, {\cal Z}_V \, V_1 \, - \, {\cal B} \, D_z K_6
\label{vmix1}
\end{equation}
and impose
\begin{equation}
\langle V_{1 \, ren} \, \bar K_{6 \, ren} \rangle \, = \, 0 \, . \label{ortVK}
\end{equation}
Here $K_{6 \, ren}$ must be as before, while we take ${\cal Z}_V,\, B$ in
standard form (\ref{renz}), (\ref{renb}) but with unknown coefficients. The
simultaneous expansion of (\ref{ortVK}) in $\epsilon$ and $g^2$ yields three
conditions from the vanishing of the constant term at $g^2$ and the pole and
the simple logarithm at $g^4$. The result is
\begin{equation}
{\cal Z}_V \, = \, 1 \, + \, g^2 \, \frac{15 \, M}{4 \, \epsilon} \, + \, 
O(g^4) \, , \qquad {\cal B} \, = \, g^2 \, \frac{15 M}{7} \, + \, g^4 \, \Bigl(
\frac{345 M^2}{56 \,
\epsilon} \, + \, b_{20} \Bigr) \, + \, O(g^6) \, . \label{vmix2}
\end{equation}
In the spin 3/spin 3 mixing $V_{1 \, ren}$ stays orthogonal to $D_z K_6$.
The requirement of finiteness of $\langle V_{1 \, ren} \, \bar V_{1 \, ren}
\rangle$ allows one to fix the $g^4$ part of ${\cal Z}_V$ which is in fact of
the form (\ref{renz}). \\

Last, $V_2$ may be renormalised multiplicatively as it does not mix.
Our main result is that $V_1, \, V_2$ have identical anomalous dimensions
\begin{equation}
\gamma_1 \, = \, \frac{15}{2} \, M \, , \qquad \gamma_2 \, = \, -
\frac{225}{16} \, M^2 \, ,
\end{equation}
i.e. the values predicted in \cite{new}.

\section{The graph calculation with ${\cal N}= 2$ superfields}

\subsection{A shopping spree in ${\cal N}=2$ land}

The $sl(2)$ sector contains the operators defined in (\ref{opdef}), though
in general with $k$ insertions of elementary fields that may all carry
covariant derivatives. Independently of the spin, the operators are in the same
$[0,k,0]$ representations of $SU(4)$ as the 1/2 BPS states without
derivatives. The definition (\ref{opdef}) is a highest weight projection in
an ${\cal N}=1$ formulation, while in ${\cal N}=2$ we may realise
representatives as chains of $k$ copies of the hypermultiplet $q^+$.
One returns to the physical scalars of the ${\cal N} = 4$ multiplet by
putting $\theta \, = \, 0$. The ${\cal N}=2$ two-point functions of composites
of matter fields involve relatively few supergraphs. The advantage is quite\
clear: at $O(g^2)$ we have only one contributing diagram and at $g^4$ there
are no more than fifteen singular graphs, which can be evaluated by standard
means since they lead to two-point integrals up to and including three-loop
level in momentum space.

Quantum calculations for the BPS operators themselves
give zero, but divergences do arise when the hypermultiplets (or $Z$'s) carry
Yang-Mills covariant derivatives. We find that the partial
derivative in ${\cal D}_\mu \, = \, \partial_\mu \, + \, i \, g \, A_\mu$ does
not introduce singularities but that the connection part gives rise to a set
of singular diagrams. This seems logical since the hypermultiplet itself
and its partial derivatives are short objects from an ${\cal N}=2$ perspective
while the connection is not.

We will not review the ${\cal N}=2$ formalism in this letter; the interested
reader is referred to \cite{HSS}. Our conventions are taken from the second
reference therein. The ${\cal N}=2$ $x$-space supergraph technique was
developed in the series of papers \cite{us1}. The ${\cal N}=4$ action is
\begin{eqnarray}
S_{N=4} & = & S_q \, + \, S_V \, + \, \text{ghost} \, + \text{gauge-fixing} \\
S_q & = & - Tr \bigl( \int du d \zeta^{-4} (\tilde q^+ \, D^{++} \, q^+ \, +
\, g \, i \, \tilde q^+ [V^{++}, q^+] ) \bigr) \label{actq} \\ S_V & = &
\frac{1} {2 g^2} \sum_{n=2}^\infty \frac{(-i g)^n}{n} \, Tr \bigl( \int
d^{12}z du_1 \ldots du_n \frac{V^{++}(z,u_1) \ldots V^{++}(z,u_n)}{(12)(23)
\ldots (n1)} \bigr) \label{actV}
\end{eqnarray}
with $\zeta = \{x_A,\theta^+,\bar\theta^+\}$ and $z = \{x,\theta,\bar\theta\}$.
The ``harmonics'' $u \in SU(2)$ are auxiliary bosonic variables. We use the
shorthand $(23) = u_2^{+i} {u_3^+}_i$, where 2,3 label distinct points and
$i \in \{1,2\}$, while the row index is split into $\{+,-\}$
according to the charge under the $U(1)$ subgroup. The $u$-variable is
used as a projector: $\theta^\pm = u^\pm_ i \theta^i$ and similarly for
$\bar \theta$ and the spinorial derivatives $D, \bar D$. The ``analytic $x$''
\begin{equation}
x_A^\mu \, = \, x^\mu - i (\theta^+ \sigma^\mu \bar \theta^- \, + \,
\theta^- \sigma^\mu \bar \theta^+)
\end{equation}
is annihilated by $D^+, \bar D^+$. In the YM action the $u$-variables
distinguish the various $V$-fields. The $SU(2)$ structure of the three-vertex
is in fact completely antisymmetric so that the simple product from
(\ref{actV}) becomes a commutator.

The two unconstrained quantum fields of the formalism are the hypermultiplet
$q^+(x_A,\theta^+,$ $\bar \theta^+,u)$ and the YM prepotential $V^{++}(x_A,
\theta^+,\bar \theta^+,u)$. We have the propagators
\begin{eqnarray}
\Pi_{12} & = & \langle q^+(1) \tilde q^+(2) \rangle \, = \, e^{\Delta_{12}}
\frac{1}{-4 \pi^2 x_{A1A2}^2} \, , \\ \Delta_{12} & = &
\frac{2i}{(12)} \bigl[ (1^-2) \theta_1^+ \partial_1 \bar \theta_1^+ +
(2^-1) \theta_2^+ \partial_1 \bar \theta_2^+ + \theta_1^+ \partial_1 \bar
\theta_2^+ + \theta_2^+ \partial_1 \bar \theta_1^+ \bigr] \nonumber \, ,
\end{eqnarray}
and in Feynman gauge
\begin{equation}
\langle V^{++}(1) V^{++}(2) \rangle \, = \, \delta(u_1,u_2)
\frac{(\theta_1^+ - \theta_2^{1+})^2 (\bar \theta_1^+ - \bar \theta_2^{1+})^2}
{2 \pi^2 x_{12}^2} \, .
\end{equation}
With the notation $\theta_2^{1+}$ we mean $\theta$ at point 2, projected by
$u^+$ from point 1. (The expression for the propagator remains reasonable due
to the presence of the $SU(2)$ delta-function.) Note that there is no higher
matter/YM interaction than the cubic term in $S_q$, while there exist YM
self-interactions of any order. Our set of graphs involves only the cubic
YM vertex, though. From the interaction Lagrangian we retain
\begin{eqnarray}
&& - g \, i \, \int du \, d\zeta^{-4} \, Tr \bigl( \tilde q^+ [V^{++}, q^+]
\bigr) \, , \\ && \frac{g \, i}{12} \, \int d^{12}z \, du_1 \, du_2 \, du_3
\frac{Tr \bigl( V^{++}(z,u_1) [V^{++}(z,u_2), V^{++}(z,u_3)] \bigr)}
{(12)(23)(31)} \, .
\end{eqnarray}
The SSDR scheme means to leave the spinor and $SU(2)$ algebra untouched but
to modify the $1/x^2$ part of the propagators as in (\ref{ssdr}). \\

Finally, we need to discuss the connection in ${\cal D}_\mu \, = \,
\partial_\mu \, + \, g \, i \, A_\mu$. In the $\lambda$-frame \cite{HSS} we
have
\begin{equation}
{\cal D}^+_\alpha \, = \, D^+_\alpha \, , \qquad
\bar {\cal D}^+_\alpha \, = \, \bar D^+_\alpha \, .
\end{equation}
(Hence the + projected spinor derivatives are YM covariant.) The - projected
spinor derivatives inherit a connection from the $SU(2)$ derivative
${\cal D}^{--}$:
\begin{eqnarray}
{\cal D}^{--} & = & D^{--} \, + \, g \, i \, V^{--} \\
{\cal D}^-_\alpha & = & [{\cal D}^{--}, D^+_\alpha] \, = \, D^-_\alpha
\, - \, g \, i \, (D^+_\alpha V^{--}) \\
\bar {\cal D}^-_{\dot \alpha} & = & [{\cal D}^{--}, \bar D^+_{\dot \alpha}]
\, = \, \bar D^-_{\dot \alpha} \, - \, g \, i \, (\bar D^+_{\dot \alpha}
V^{--})
\end{eqnarray}
The super-algebra does not change under covariantisation:
\begin{equation}
\{ {\cal D}^+_\alpha, \bar {\cal D}^-_{\dot \alpha} \} \, = \, - 2 i \,
{\cal D}_{\alpha \dot \alpha}
\end{equation}
from where we may read off
\begin{equation}
{\cal D}_\mu \, = \, \partial_\mu \, + \, \frac{g}{4} (D^+ \sigma_\mu \bar D^+
\, V^{--}) \, . \label{connect}
\end{equation}
The field $V^{--}$ has a series expansion in terms of $V^{++}$
quite like the YM action:
\begin{eqnarray}
V^{--}(z,u_1) & = & \int du_2 \, \frac{V^{++}(z,u_2)}{(12)^2} \, + \, \\
& + & \frac{g i}{2} \, \int du_2 \, du_3 \, \frac{[V^{++}(z,u_2), V^{++}
(z,u_3)]}{(12)(23)(31)} \, + \, \ldots
\end{eqnarray}
We draw the reader's attention to the non-linearity of the connection
(\ref{connect}) in $V^{++}$. For our purposes only the first two terms of the
expansion are needed. Note that the linear term has the following two-point
function with $V^{++}$:
\begin{equation}
\langle \, \frac{1}{4} (D^+ \sigma_\mu \bar D^+
\, V^{--})(1) \, V^{++}(2) \, \rangle_{\theta_1, \bar \theta_1 = 0} \, = \, -
\, e^{\frac{2i}{(12)} \, (2^-1) \theta_2^+ \partial_1 \bar \theta_2^+} \,
\frac{\theta_2^+ \sigma_\mu \bar \theta_2^+}{2 \pi^2 (x_1 - x_{A2})^2}
\end{equation}
When evaluating Feynman graphs with connection ends it is a very
useful fact that the exponential shift in this formula coincides
with one of the terms in the shift acting on the matter
propagator. This exponential is usually reproduced also in graphs
involving the quadratic part in the connection because cross terms
from the denominators of the attached propagators are eliminated
when putting the outer $\theta$'s to zero.

Let us comment on how to evaluate supergraphs. A YM-line between two
matter/YM vertices is relatively easy to deal with: owing to the presence of
the $u$ delta-function and the nilpotent numerator we may change the
denominator of the $V$-propagator from $x_1 - x_2$ to $x_{A1} - x_{A2}$. One
$SU(2)$ integration can trivially be done using the delta-function. We may now
use the numerator of the propagator as a delta-function for the + projection
of the two $\theta$'s, the minus projection is absent. Thus we
replace, say the spinors in the exponential shifts $\Delta$ at the second
vertex by those from the first and integrate away the delta-function. The
second $\theta$ and $u$ integration is left over for the moment.

Graphs involving the cubic YM vertex or a connection end are trickier in that
a given $\theta$  may occur in contractions w.r.t. several different $u$'s.
If the YM lines end on matter lines it can be an advantage to shift to a chiral
basis at the complicated end, otherwise it is hard to give general rules.
In order to integrate out the $\theta$ at the cubic vertex it is always
advisable to choose that projector which occurs most often and to use the
``cyclic identity''\footnote{For three harmonics: $(12) u_3 + (23) u_1 + 
(31) u_2 \, = \, 0$, similarly with one or several - projections.} to write all
occurrences of the $\theta$ in question in terms of the $+,-$ projections
w.r.t. this one harmonic.

Eventually one has to expand the exponentials and the numerators in the
remaining $\theta$'s, leading to a linear combination of differential
operators under the $x$-integrations with harmonic (i.e. $SU(2)$) integrals
as coefficients. These $u$-integrations can always be done in an algebraic
fashion: if there is no denominator one can straightforwardly express them
in terms of $SU(2)$ epsilon tensors, else one pulls a $D^{++}$ off the
numerator and partially integrates it onto the denominator of the integral
thus producing delta-functions \cite{HSS}. For the more complex $g^4$
supergraphs we admittedly had to resort to a \emph{Mathematica} programme to
do parts of the Grassmann algebra and the $SU(2)$ integrations.

In the pictures below we have split the lines emanating from the outer
points. We do so purely for convenience of drawing --- all lines coming in from
the left start at $z_1 \,=\, \{x_1, \theta_1, u_1\}$ and all lines ending on
the right come together at the point with coordinates $z_2 \, = \,\{x_2,
\theta_2,u_2\}$. ``Spectator lines'',~i.e. free matter propagators have been
omitted. There are, of course, as many as needed to connect all matter fields
in the composite operators at points $z_1, \, z_2$. We did not indicate the
outer partial derivatives either since they do not interfere with the
Grassmann- and $SU(2)$-integrations.

Note that any two-point graph related to a $[0,k,0]$ operator will come with
a factor $(12)^k$, which we simply drop. It is equal in the tree-level and the
$g^2$ and $g^4$ parts and universal to all operators built from $k$ fields;
therefore it does not influence the renormalisation at all.
At order $g^4$ individual graphs also produce ``harmonic-nonanalytic'' terms
coming with $(12)^{k+1} (1^-2^-)$, which must eventually cancel. A good sign
is that all such terms contain free propagators times the finite integral $f$
from graph $B_0$ below. We did not explicitly check the vanishing of their sum
because they are irrelevant to the second anomalous dimension.

\subsection{Graphs}

At order $g^2$ there are only two graphs, for which we find the $x$-space
expressions
\begin{eqnarray}
B_0 & = & - (12)^2 \, \square_1 \, f(1,2;1,2) \, ,
\label{graphB0} \\ G_0 & = & - (12) \phantom{^2} \, i \, \partial_2^\mu \,
g(1,1,2) \, . \label{graphG0}
\end{eqnarray}
Here we have omitted the normalisation constants of the propagators,~i.e.
$c_0 = - 1/(4 \pi^2)$ from the hypermultiplet propagator and $-2 c_0$ from
the normalisation of the $V$-line. The integrals $f,g$ are as the pictures
suggest --- there is no extra numerator and all lines are to be replaced by
$1/x_{ij}^2$, hence
\begin{equation}
f(1,2;1,2) \, = \, \int \frac{d^4x_3 d^4x_4}{x_{13}^2 x_{14}^2 x_{34}^2
x_{23}^2 x_{24}^2} \, , \qquad g(1,1,2) \, = \, \int \frac{d^4x_3}{(x_{13}^2)^2
x_{23}^2} \, . \label{intex}
\end{equation}
The topology of a graph cannot change due to the spinor and $SU(2)$ algebra
other than by going into a ``derived topology'',~i.e. one in which some lines
are shrunk to points by box operators.

\vskip 0.1 in

\begin{minipage}{\textwidth}
\begin{center}
\includegraphics[width=0.60\textwidth]{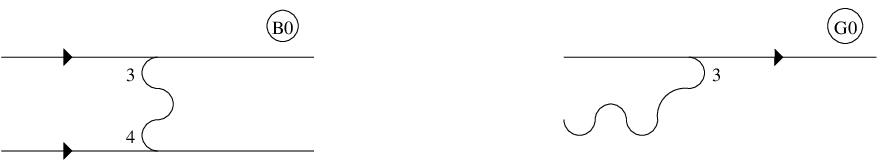}
\end{center}
\end{minipage}
\begin{center}
Figure 1. Graphs at order $g^2$
\end{center}

Let us focus on diagram $B_0$. The double-box integral $f$ is finite and
by dimensionality we have $f(1,2;1,2) \propto 1/x_{12}^2$, so that the outer
box operator produces a delta-function. In the SSDR scheme this is seen as
an $O(\epsilon)$ term, see Section 2. This would be the only contribution in
a two-point function of products of hypermultiplets without any derivatives.
Indeed, such operators are 1/2 BPS and their two-point functions should not
receive any quantum corrections except for contact terms.

What happens when the diagram has partial derivatives on the outer legs?
The \emph{Mincer} system can calculate $f$ in momentum space, and it can also
handle arbitrary \emph{scalar} numerators. We use the programme according
to the following strategy:
\begin{itemize}
\item All partial derivatives are translated into momentum space as momenta
with open indices. We make the sets of indices symmetric and traceless
separately at the left and at the right end of the
graph, so that they define spin $s_l$ and spin $s_r$ ``harmonic tensors'',
respectively.
\item There are $Min(s_l,s_r)+1$ independent ways of projecting this with
the total ingoing momentum $q$, because we may choose to contract 0 through
$Min(s_l,s_r)$ indices with the metric $\eta$. All remaining open indices are
contracted on $q$.
\item For any given distribution of outer partial derivatives/momenta we
calculate all such projections by \emph{Mincer}.
\item The integral with open indices (symmetrised and traceless at each end)
contains exactly as many independent Lorentz structures as there are
projections, the criterion being once again how many $\eta$ symbols with
one index from the left set and one from the right set are involved. It is a
simple exercise to set up and invert the set of linear equations which
allows to reconstruct the coefficient of each basis element from the
projections. The coefficients are generally $\epsilon$-dependent; we keep
$O(1), O(\epsilon)$.
\item We Fourier-transform back to $x$-space.
\end{itemize}
In this context, too, it is actually most convenient to discard all traces
since under the FT
\begin{equation}
(i q)^{(s_l+s_r)} (q^2)^{-2 \epsilon} \, \rightarrow \, \frac{2 \epsilon
(1 - 3 \epsilon + \ldots)}{\pi^2} \, (-\partial_1)^{(s_l+s_r)} \,
\frac{1}{(x^2_{12})^{(2-3 \epsilon)}} \label{fourier1l}
\end{equation}
and the term without $\eta$ is simply obtained by having all derivatives act
on the denominator. \\

For $B_0$ we have scanned all spins up to three derivatives at one end and four
at the other. Our findings are surprising: the term without $\eta$ remains a
contact term for any harmonic tensors at the left and at the right end, while
terms with $\eta$ may be finite. In conclusion, the ``BPS-like'' diagram
$B_0$ drops out of the calculation of the first anomalous dimension.

The calculation of $G_0$ can be done in a variety of ways because the
underlying integral $g(1,1,2)$ is elementary. Since our computer
programmes were originally geared at the $[0,2,0]$ operators we have chosen to
join one free line and compute the resulting two-loop diagram as a derived
topology of the T2 graph \cite{Mincer}, alternatively one may use T1. The
results of \emph{Mincer} have to be scaled down by a factor $(4 \pi)^{2l}$
where $l$ is the loop order of the given integral. In configuration space
graph $G_0$ has a simple pole.\footnote{The index of the connection
``migrates'' to the opposite end of the graph, see (\ref{graphG0}).
When building up the harmonic tensors we must put this index into the tensor
at the left end, of course.}

We remark that the diagrams have to be recomputed for every different
distribution of derivatives. The combinatorics becomes more elaborate, too,
since equal looking legs can be distinguished by the number of partial
derivatives.

On the computing side, we have implemented the large $N$
combinatorics for the $[0,3,0]$ operators in a
$\emph{Mathematica}$ programme, which uses a table of integrals to
calculate the matrices of two point functions at tree-level and
orders $g^2,g^4$. The calculation of each integral requires the
construction of the necessary harmonic tensors and their projections; for
convenience of programming we still do this under
\emph{Mathematica}. Finally, we link in \emph{Mincer} under
\emph{Form} to evaluate the projected integrals. \\

Let us switch to order $g^4$. There are five ``BPS-like'' graphs,~i.e.
topologies that one also finds in two-point functions of 1/2 BPS
operators, though without extra partial derivatives on the outer legs.
When analysing two-point functions of 1/2 BPS operators one tends to find two
linear combinations of graphs coming with independent colour structures
\cite{contact}. Protectedness is tantamount to both of these linear
combinations being contact terms. In the manifestly finite setup we use
(${\cal N}=4$ in terms of ${\cal N}=2$ fields) the only divergences are
associated to composite operators. Therefore, graphs related to BPS objects
should even be individually finite.

The BPS-like supergraphs are drawn in Figure 2. In $B_2$ there is no vertex in
the middle of the diagram, likewise in $B_5$ the YM line between points 5 and
6 does not touch the matter line in the middle of the diagram.

\vskip 0.2 in

\begin{minipage}{\textwidth}
\begin{center}
\includegraphics[width=0.80\textwidth]{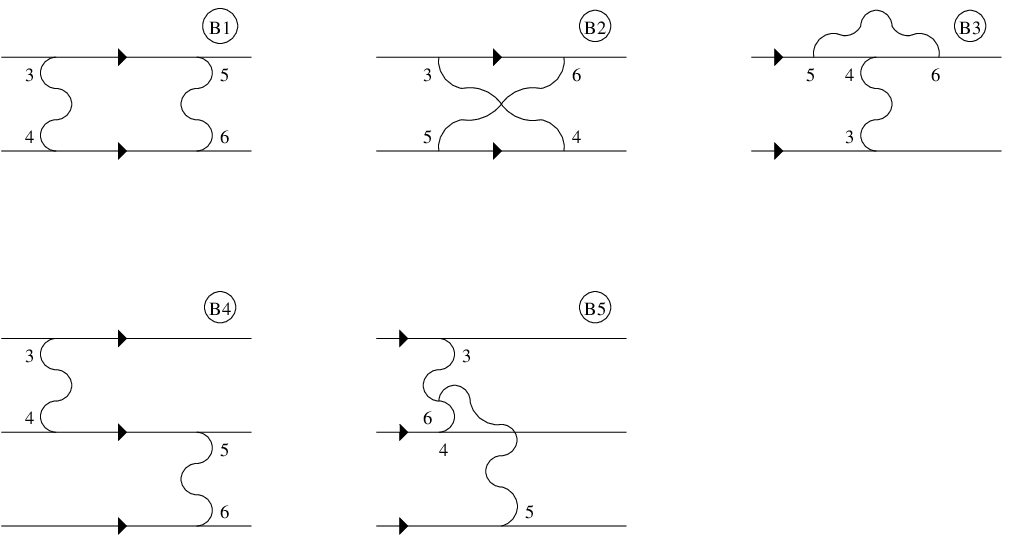}
\end{center}
\end{minipage}
\vskip 0.1 in
\begin{center}
Figure 2. BPS-like graphs at order $g^4$.
\end{center}

\noindent We find
\begin{eqnarray}
B_1 & = & (12)^2 \, \square_1 \square_1 \, LA \, , \label{intB1} \\
B_2 & = & (12)^2 \, \bigl[ - \square_1 (\partial_{13} +
\partial_{24})^2 + (12)(1^-2^-) (\square_{15} \square_{24} +
\square_{13} \square_{26}) \bigr] \, NP \, , \label{intB2} \\
B_3 & = & (12)^2 \, \bigl[ - \square_1 \square_{34} \, + \,
(12)(1^-2^-) (\square_{15} \square_{23} + \square_{13} \square_{26}) \bigr] \,
BE \, , \label{intB3} \\ B_4 & = & (12)^3 \, \bigl[(\partial_{13} +
\partial_{14})^2 (\partial_{25} + \partial_{26})^2 -
\square_1 \square_{45} / 2 - \square_{16} \square_{23} + \label{intB4} \\ &&
\phantom{(12)^3 \, \bigl[} + (12)(1^-2^-) \square_{16}
\square_{23} \bigr] \, P4 \nonumber \, ,
\end{eqnarray}
where the normalisation constants of the propagators were again omitted.
The $x$-space integrals $LA,\,NP,\,BE$ are the so-called ``ladder'',
``non-planar'' and ``Benz'' three-loop integrals as depicted in Figure 2.
We illustrated by equation (\ref{intex}) how to read them off from the
graphics; there is no additional numerator and the denominator is given by
the appropriate product of $1/x_{ij}^2$, no propagator is doubled.
By $\partial_{ij},\square_{ij}$ we
mean derivatives acting at point $i$ on a line going from $i$ to $j$, whereas
$\square_1$ means an outer box-operator acting on the whole expression.

All three topologies $LA,\,NP$ and $BE$ are finite without a numerator, but
differentiation under the integral could potentially make them divergent.
Using the \emph{Mincer} system we have checked that the complete expressions
$B_1,\,B_2,\,B_3$ do not diverge either. Like $B_0$ the graphs remain finite,
too, when symmetrised, traceless products of derivatives are introduced acting
from the left and from the right, respectively. A scan found no singularities
up to spin 3 at one end and spin 4 at the other.

The integral in graph $B_4$ was baptised ``P4'' above as it is planar and has
four loops. A quick evaluation of (\ref{intB4}) in point-splitting suggests
that the graph without extra derivatives is finite. Unfortunately, we
have no way of calculating $B_4$ in dimensional regularisation, in particular
in the presence of outer partial derivatives. By analogy with $B_0 \ldots B_3$
we assume that $B_4$ is finite, too, so that we may dispose of it. 
Diagram $B_5$ should behave accordingly; at any rate, as a non-planar
structure it is not relevant in the context of our leading $N$ analysis.

Note that the non-analytic parts in (\ref{intB2}) ... (\ref{intB4}) always
reduce to $f(1,2;1,2)$ whence they are finite by what was said about $B_0$.

The singular graphs at order $g^4$ are listed in Figure 3. We have omitted the
mirror graphs and some diagrams that must carry at least one power of $\eta$
on dimensional grounds, e.g. the graph with three connection lines joining at
a cubic YM vertex. The product $B_0 * G_0$ has no singular term without $\eta$
either, see above.

The graphs in the third row carry an index $\mu$ at point 1 on the left and an
index $\nu$ at point 2 on the right. The dot in $G_4,\,G_7,\,G_8$ indicates
the quadratic term in the connection from which two $V$-lines emanate, whereas
$G_{13},\,G_{14},\,G_{15}$ have two different occurrences of the connection on
the left with the indices $\mu,\nu$.

\vskip 0.25 in

\begin{minipage}{\textwidth}
\begin{center}
\includegraphics[width=0.80\textwidth]{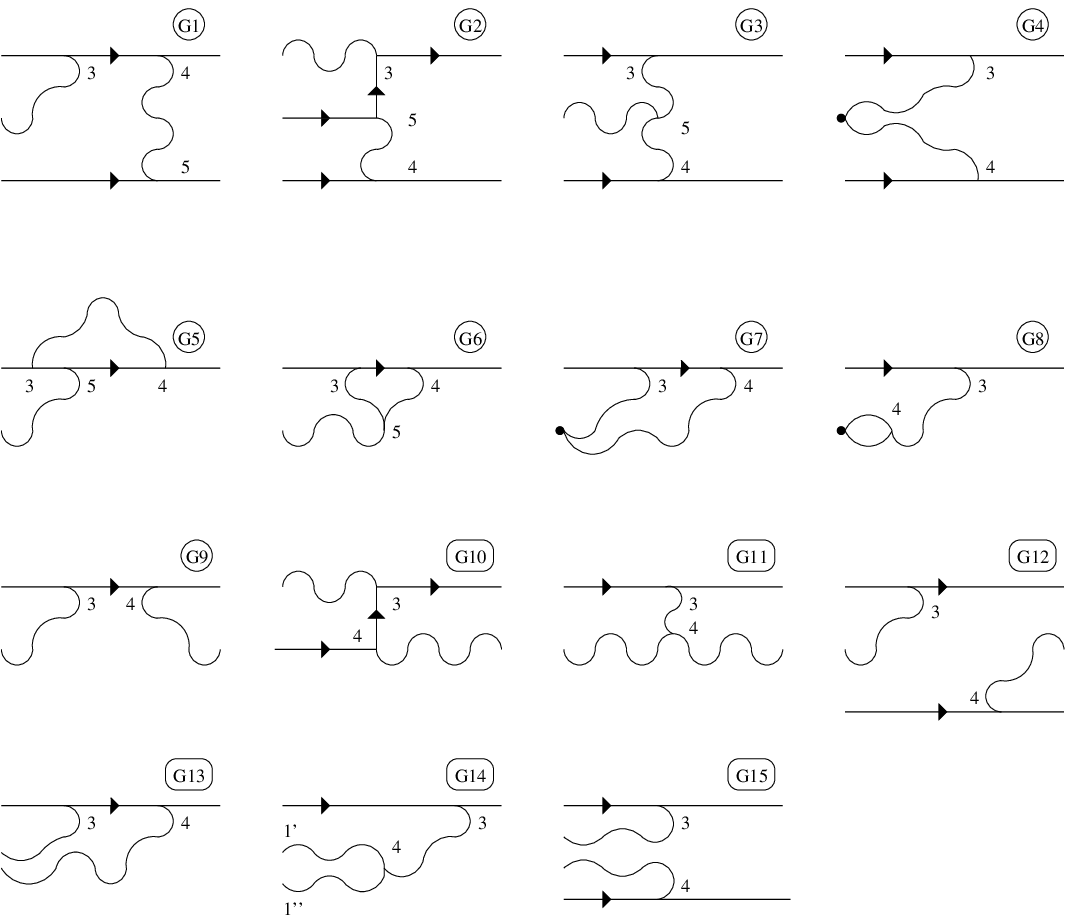}
\end{center}
\end{minipage}
\begin{center}
Figure 3. $O(g^4)$ graphs with connection lines.
\end{center}

Evaluation of the Grassmann- and $SU(2)$-integrations reduces the supergraphs
to the following differential operators under the appropriate integrals:
\begin{eqnarray}
G_1 & = & (12)^2 i \bigl[ \, \phantom{-} \; \partial_{\mu 1} \square_{34} / 2 -
\partial_{\mu 34} \square_1 \, \bigr] \label{intG1} \\
G_2 & = & (12)^2 i \bigl[ \, \phantom{-} \; \partial_{\mu 1} \square_{35} / 2 +
\partial_{\mu 23} (\partial_{14} + \partial_{15})^2 \, \bigr] \label{intG2} \\
G_3 & = & (12)^2 i \bigl[ \, \phantom{-} \; \partial_{\mu 1}
\square_{35}/2 + \partial_{\mu 23} (\partial_{14} + \partial_{15})^2 -
\partial_{\mu 23} \square_{14} / 2 \label{intG3} \\
 && \phantom{(12)^2 i \bigl[} 
- \partial_{\mu 1} \square_{45} / 2 - \partial_{\mu 24} (\partial_{13} +
\partial_{15})^2 + \partial_{\mu 24} \square_{13}/2 \nonumber \\
&& \phantom{(12)^2 i \bigl[} -(\partial_{\mu 23} - \partial_{\mu 24})
\square_{15} / 2 \nonumber \\
&& \phantom{(12)^2 i \bigl[} -(\partial_{\mu 23} - \partial_{\mu 24})
\square_1 / 2 \nonumber \\
&& \phantom{(12)^2 i \bigl[} - \partial_{\mu 13} (\square_1 +
\square_{23} - 2 \square_{24}) / 2 + \partial_{\mu 14} (\square_1 -
2 \square_{23} + \square_{24}) / 2 \nonumber \\
&& \phantom{(12)^2 i \bigl[} + \partial_{\mu 1} (\square_{23} -
\square_{24}) - (12)(1^-2^-) \partial_{\mu 15} (\square_{23} - \square_{24})
\, \bigr] \nonumber
\end{eqnarray}
\begin{eqnarray}
G_4 & = & (12)^2 i \bigl[ \, - \; (\partial_{\mu 23} - \partial_{\mu 24})
/ 2 \, \bigr] \label{intG4} \\
G_5 & = & (12) \phantom{^2} i \bigl[ \, - \; \partial_{\mu 2}
\square_{15}  - (12)(1^-2^-) \partial_{\mu 15} \square_1 \, \bigr]
\label{intG5} \\
G_6 & = & (12) \phantom{^2} i \bigl[ \, - \; \partial_{\mu 2}
\square_{15} + \partial_{\mu 2} \square_{13} / 2 - \partial_{\mu 1}
\square_1 + (12)(1^-2^-) \partial_{\mu 15} \square_1 \, \bigr] \label{intG6} \\
G_7 & = & (12) \phantom{^2} i \bigl[ \, \phantom{-} \,
\partial_{\mu 2} \, \bigr] \label{intG7} \\
G_8 & = & (12) \phantom{^2} i \bigl[ \, \phantom{-} \,
\partial_{\mu 2} \, \bigr] \label{intG8} \\
G_9 & = & (12) \phantom{^2 i} \bigl[ \, - \, \partial_{\mu 34}
\partial_{\nu 34} + \eta_{\mu \nu} \square_{34} / 4 \, \bigr] \label{intG9} \\
G_{10} & = & (12) \phantom{^2 i} \bigl[ \, \phantom{-} \,
\partial_{\nu 14} \partial_{\mu 23} +  \partial_{\mu 34} \partial_{\nu 34} -
\eta_{\mu \nu} \square_{34} / 4 \label{intG10} \\ && \phantom{(12) ^2 i}
\, \, - (1 + (12)(1^-2^-)) \partial_{\mu 13} \partial_{\nu 24} \, \bigr]
\nonumber \\
G_{11} & = & (12) \phantom{^2 i} \bigl[ \, - \, \partial_{\nu 13}
\partial_{\mu 23} + \partial_{\mu 1} \partial_{\nu 2} - (\partial_{\mu 14}
\partial_{\nu 23} + \partial_{\mu 13} \partial_{\nu 24}) / 2 \label{intG11} \\
&& \phantom{(12)^2 i \bigl[ } + \eta_{\mu \nu} (\partial_{13} -
\partial_{23}). (\partial_{14} - \partial_{24}) / 4 -
(12)(1^-2^-) \partial_{\mu 14} \partial_{\nu 24} \, \bigr]  \nonumber \\
G_{12} & = & (12)^2 \phantom{i} \bigr[ \, \phantom{-} \, \partial_{\mu 23}
\partial_{\nu 14} \, \bigr] \label{intG12} \\
G_{13} & = & (12) \phantom{^2 i} \bigr[ \, \phantom{-} \,
\partial_{\mu 34} \partial_{\nu 2} + \eta_{\mu \nu} \square_{34} / 4 \, \bigr]
\label{intG13} \\
G_{14} & = & (12) \phantom{^2 i} \bigl[ \, \phantom{-} \,
\partial_{\mu 23} \partial_{\nu 14}' - \partial_{\nu 23} \partial_{\mu 14}''
- (\partial_{\nu 23} \partial_{\mu 14}' - \partial_{\mu 23}
\partial_{\nu 14}'') / 2 \label{intG14} \\
&& \phantom{(12)^2 i \bigl[} + \eta_{\mu \nu} (\partial_{13} -
\partial_{23}).(\partial_{14}' - \partial_{14}'') / 4 \, \bigr]  \nonumber \\
G_{15} & = & (12)^2 \phantom{I} \bigl[ \, - \, \partial_{\mu 23}
\partial_{\nu 24} \, \bigr] \label{intG15}
\end{eqnarray}
The combinatorics for $G_4$ is like for $G_3$ because the quadratic term in
the connection is a commutator. The graphs come in a fixed sum
in which $G_4$ compensates the $\square_{15}$ term in the expression
for $G_3$. The box operators $\square_{23},\square_{24}$ in (\ref{intG3})
reduce the underlying $BU$ topology to the usual finite $f(1,2;1,2)$.
In all these terms the remaining derivative acts from the left and thus
there are no poles.
The $BU$ integral itself is in fact finite, too, as long as no double lines
are introduced. We have checked that the $\square_1$ terms in the fifth line of
(\ref{intG3}) may be dropped as well. The integral is slightly more
robust than $f$ in that the term $(\partial_{\mu 23} - \partial_{\mu 24})
\square_1$ is not singular either, although the derivatives act from the right.
In conclusion,
\begin{eqnarray}
G_3 + G_4 & = & (12)^2 i \bigl[ \, \phantom{-} \; \partial_{\mu 1}
\square_{35}/2 + \partial_{\mu 23} (\partial_{14} + \partial_{15})^2 -
\partial_{\mu 23} \square_{14} / 2 \label{g3g4} \\
 && \phantom{ (12)^2 i \bigl[} 
- \partial_{\mu 1} \square_{45} / 2 - \partial_{\mu 24} (\partial_{13} +
\partial_{15})^2 + \partial_{\mu 24} \square_{13}/2 + \ldots \, \bigr]
 \nonumber
\end{eqnarray}
where the dots indicate the omitted finite terms. Similarly, $G_5,\,G_6,\,G_7$
come in a fixed sum. On discarding finite terms:
\begin{equation}
G_5 + G_6 + G_7 \, = \, (12) \, i \bigl[ \, - \, \partial_{\mu 2}
\square_{15} + \partial_{\mu 2} \square_{13} / 2 + \ldots \bigr]
\end{equation}
(The numbering of the points refers to $G_5,G_6$.) Owing to the box operators
both terms can be fitted into the $O2$ topology after joining a free line.
It is interesting to note that the $\square_{13},\square_{14}$ terms in
(\ref{g3g4}) are of the same type, while the remaining terms are as in $G_2$.
The graphs $G_1 \ldots G_7$ seemingly always sum to zero if one
end of the correlator is a BPS state. Diagram $G_8$ has the same
combinatorics as the other graphs in the second line, but it is a separate
piece in this sense.

In diagrams $G_{10},G_{11}$ the potentially divergent pieces without $\eta$
are
\begin{eqnarray}
G_{10} & = & (12) \phantom{^2} \bigl[ \, \phantom{-} \,
\partial_{\nu 14} \partial_{\mu 23} +  \partial_{\mu 34} \partial_{\nu 34} 
+\ldots \, \bigr] \, , \\
G_{11} & = & (12) \phantom{^2} \bigl[ \, - \, \partial_{\nu 13}
\partial_{\mu 23} + \ldots \, \bigr] \, .
\end{eqnarray}
In our calculations the complete sum of graphs turned out to be finite within
each class.

One last simplification concerns graph $G_{14}$: for our purposes we may once
again drop the $\eta$ term. Further, when the two indices are symmetrised
\begin{equation}
G_{14} \, = \, (12) \phantom{^2} \bigl[ \,
\partial_{(\mu 23} \bigl(\partial_{\nu) 14}' - \partial_{\nu) 14}''\bigr) / 2 
+ \ldots \, \bigr] \, .
\end{equation}
We remark that diagram $G_{14}$ starts to contribute only at spin 4.

All the integrals can be evaluated by the \emph{Mincer} system just as
the order $g^2$ examples: where there is only one matter line we join a second,
free one. In this way $G_1,\,G_5 \ldots G_9$, $G_{12} \ldots G_{14}$ are of
topology $O2$ (or derived of $O2$) while the remaining diagrams belong
to the $BU$ category. The rest of the algorithm is as before, although
formula (\ref{fourier1l}) must be adapted to the dimension of the three-loop
integrals:
\begin{equation}
(i q)^{(s_l+s_r)} (q^2)^{-3 \epsilon} \, \rightarrow \, \frac{3 \epsilon
(1 - 4 \epsilon + \ldots)}{\pi^2} \, (-\partial_1)^{(s_l+s_r)} \,
\frac{1}{(x^2_{12})^{(2-4 \epsilon)}} \label{fourier2l}
\end{equation}

\section{Conclusions and outlook}

We have discussed the descendant structure and renormalisation of the $sl(2)$
sector operators of length three up to spin three. These are a 1/2 BPS state
$O$, a spin two operator $K_6$ belonging to the second lowest two-impurity BMN
multiplet and two spin three primaries $V_1, \, V_2$ with identical
scaling dimensions up to order $g^4$. The anomalous dimensions are
\begin{equation}
K_6 \, : \, \gamma_1 \, = \, 4 M, \, \gamma_2 \, = \,
- 6 M^2 \, , \qquad V_1, \, V_2 \, : \, \gamma_1 \, = \, 
\frac{15}{2} M, \, \gamma_2 \, = \, - \frac{225}{16} M^2  \nonumber
\end{equation}
with $M \, = \, N / (4 \pi^2)$. The result for $V_1, \, V_2$ is in exact
agreement with \cite{new},~c.f. Table 2.

The most difficult step in developing any higher-loop dilatation operator is
to make it work when the spin-chain length is smaller than the maximum number
of adjacent sites on which the operator can act (``wrapping''). From this
perspective, the hardest test on any universal formula is to go from the
shortest spin-chain length to operators with subsequently more sites in the
trace. In the $sl(2)$ sector this would be the step from twist two to twist
three, which we have checked for the Bethe-ansatz \cite{new}, initially at the
lowest non-trivial spin.

The obvious continuation of our work is to investigate other mixing examples
at higher spin as well as for operators with more elementary fields. The
ultimate aim of the line of research is to set up a two-loop dilatation
operator for the $sl(2)$ sector. It would be interesting to see whether the
dilatation operator can be directly constructed from the elements of our
renormalisation procedure by going back to the non-orthogonal basis
$\{s_1,s_2,s_3\}$ defined in equation (\ref{opdef}).

The renormalisation of $V_1$ is non-multiplicative: we have put
$V_{1 \, ren} \, = \, {\cal Z}_V \, V_1 \, - \, {\cal B} \, D_z K_6$ with a
singular factor ${\cal B}$ that cannot be split into $b(g^2) \, {\cal Z}_V$,
where $b(g^2)$ is some finite function. This example of orthogonalisation in
dimensional regularisation closely parallels the discussion of the Konishi
anomaly in \cite{Eden:2005ve}.

Last, the limiting step in our current software is the manipulation of harmonic
tensors, because for higher spin these involve very many terms. It would
clearly be desirable to transfer a larger part of the calculation from
\emph{Mathematica} to \emph{Form}.

\section*{Acknowledgements}

The author benefited from many useful discussions with G. Arutyunov,
N. Beisert, V. Dippel, E. Sokatchev and M. Staudacher. He would also like
to thank C. Jarczak for start up hints on xfig and G. Arutyunov, E. Sokatchev
and M. Staudacher for comments on the manuscript.

\end{document}